\documentclass[aps,prl,twocolumn,superscriptaddress,showpacs]{revtex4-1}
\usepackage{amsmath}
\usepackage{graphicx,graphics}
\usepackage{hyperref}
\usepackage{amssymb}
\usepackage{color}
\usepackage{appendix}
\begin{document}

\title{BCS-BEC crossover in a quasi-two-dimensional Fermi superfluid}
\author{Jing Zhou}
\thanks{These authors contributed equally to this work.}
\affiliation{Department of Science, Chongqing University of Posts and Telecommunications,
Chongqing 400065, China}
\author{Tingting Shi}
\thanks{These authors contributed equally to this work.}
\affiliation{Department of Physics, Renmin University of China, Beijing 100872, China}
\author{Xia-Ji Liu}
\affiliation{Centre for Quantum Technology Theory, Swinburne University of Technology,
Melbourne, Victoria 3122, Australia}
\author{Hui Hu}
\thanks{corresponding author:hhu@swin.edu.au}
\affiliation{Centre for Quantum Technology Theory, Swinburne University of Technology,
Melbourne, Victoria 3122, Australia}
\author{Wei Zhang}
\thanks{corresponding author:wzhangl@ruc.edu.cn}
\affiliation{Department of Physics, Renmin University of China, Beijing 100872, China}
\address{Beijing Academy of Quantum Information Sciences, Beijing 100193, China}
\date{\today }
\begin{abstract}
We study the crossover from the Bardeen-Cooper-Shrieffer (BCS) regime to the Bose-Einstein-condensation (BEC) regime in a quasi-two-dimensional quantum gas of ultracold fermionic atoms. Using an effective two-dimensional Hamiltonian with renormalized interactions between atoms and dressed molecules within a Gaussian pair fluctuation theory, we investigate how Fermi superfluidity is affected by reduced dimensionality at zero temperature in a wide range of crossover. We observe that the order parameter and pair size show universal relations with the chemical potential on the BCS side, irrespective of dimensionality. However, such universal dependences break down towards the BEC limit with increasing interaction strength. This results reveal the notable effect of reduced dimenionality on pairing physics, which can also be observed in the sound velocity and convexity parameter of the Goldstone mode. We compare our results with the latest experiments in both $^{6}{\rm Li}$ atomic gases and layered nitrides Li$_x$ZrNCl and find good agreements.
\end{abstract}
\maketitle

\textit{Introduction.--} The effect of reduced dimensionality on fermionic pairing is an intriguing yet open question in studies of superfluids and superconductors. Owing to the restriction of the motional degrees of freedom and the existence of strong fluctuations in low dimensions, exotic pairing phases might be stabilized, such as the Fulde-Ferrell-Larkin-Ovchinnikov state with finite pairing momentum~\cite{FFLOreview} and the Berezinskii-Kosterlitz-Thouless (BKT) phase with vortex-antivortex pairs~\cite{Berezinskii,Kosterlitz,Nelson}. Of particular interest is two-dimensional (2D) Fermi systems, which not only give the highest critical temperature at ambient pressure~\cite{tcrecord}, but also feature quantum anomaly in their collective excitations~\cite{Hofmann, Taylor, Gao, Hu1}.

Recent rapid experimental progresses have greatly promoted the research on 2D superfluidity and superconductivity. In condensed matter systems, precise control of carrier density, geometric strain and lattice superstructures significantly bring forward the study of multiband layered superconductors such as FeSe~\cite{Hanaguri-19}, FeSe$_{1-x}$S$_x$~\cite{Hashimoto-20, Mizukami-21}, Li$_x$ZrNCl~\cite{Nakagawa-21}, and magic-angle twisted trilayer graphene~\cite{Park-21}. In atomic systems, highly anisotropic traps or one-dimensional optical lattices have been implemented to realize quasi-two-dimensional (Q2D) Fermi gases with tunable interparticle interaction via Feshbach resonances~\cite{Frohlich-11, Orel-11, Sommer-12, Makhalov-14, Ries-15, Cheng-16, Fenech-16, Mitra-16, Moritz2, Moritz1}. Two latest experiments in gate-controlled layered nitrides~\cite{Nakagawa-21} and ultracold gases of $^6$Li atoms~\cite{Moritz1, Moritz2} have now provided benchmark results on 2D superfluidity by evolving the system from a weakly interacting Bardeen--Cooper-Schrieffer (BCS) superfluid to a tightly-bound Bose-Einstein condensate (BEC)~\cite{strinati-review}. In particular, in the latter experiment the superfluid gap has been suggested to follow a universal function of the interaction strength regardless of dimensionality~\cite{Moritz1}. This is an interesting hypothesis that requires timely theoretical clarification.

In this Letter, we theoretically investigate the BCS--BEC crossover of a Fermi gas subjected to a tight harmonic confinement along one spatial dimension ($z$), under conditions compatible with the recent experiments of $^6$Li atoms~\cite{Moritz1, Moritz2}. In the BCS regime, the confinement provides the dominant energy scale and the system approaches the 2D limit. In the BEC regime, the binding energy $E_B$ of the two-body bound state exceeds the confinement such that the excited harmonic oscillator states along the $z$-direction are significantly populated~\cite{Levinsen}, and the 2D regime may become difficult to reach~\cite{Wu-20}. To account for such a Q2D configuration, we adopt an effective two-channel model with dressed molecule~\cite{Duan1,Duan2,Duan3}, and use a Gaussian pair fluctuation (GPF)~\cite{NSR, SadeMelo-93, Hu2, Hu3, He-15} approach to study the pairing properties at zero temperature. Compared with the GPF results of pure 2D and 3D systems, we find that the relations between the superfluid gap $\Delta$ and chemical potential $\mu$ are only universal in the BCS regime (i.e., $\Delta / E_F \lesssim 0.5$). With increasing interaction strength, the difference between Q2D and 3D systems becomes significant, indicating a strong influence of dimensionality. The remarkable reduced dimensionality effect can also be drawn from the results of pair size, sound velocity, and convexity parameter of the Goldstone mode. Our calculations show reasonable agreement with experiments in both $^6$Li atomic gases~\cite{Moritz1, Moritz2} and layered nitrides~\cite{Nakagawa-21}.

\textit{Effective Hamiltonian and GPF theory.--}
We consider an ultracold gas of fermionic atoms confined in an axially symmetric anisotropic harmonic trap with trapping frequency along the axial direction ($\omega_z$) much stronger than that in the radial plane. When the reduced chemical potential $\widetilde{\mu}=\mu+E_{B}/2$ and the thermal energy $k_BT$ are both much smaller than the energy scale $\hbar \omega_z$, the low-energy/long-range physics of the system reduces effectively to 2D. Meanwhile, by tuning the inter-atomic interaction into the BEC regime, the excited harmonic oscillator states of the tightly confined direction are inevitably populated~\cite{Duan1}. Therefore, the high-energy/short-range details must be renormalized by either an energy dependent scattering length~\cite{Hu1,Wu-20} or a phenomenological degree of freedom of dressed molecule~\cite{Duan2, Duan3}. In the following, we take the latter approach and write the effective 2D Hamiltonian in the form of a two-channel model (with natural units $\hbar=k_{B}=1$),
\begin{eqnarray}
H_{\rm eff} & = & \sum_{\mathbf{k}\sigma}\xi_{\mathbf{k}}a_{\mathbf{k}\sigma}^{\dagger}a_{\mathbf{k}\sigma}+\sum_{\mathbf{q}}(\xi_{\mathbf{q}}+\delta_{b})d_{\mathbf{q}}^{\dagger}d_{\mathbf{q}}\nonumber\\
& &+V_{b}\sum_{\mathbf{kk'q}}a_{\mathbf{k}+\mathbf{q}/2,\uparrow}^{\dagger}a_{-\mathbf{k}+\mathbf{q}/2,\downarrow}^{\dagger}a_{-\mathbf{k}'+\mathbf{q}/2,\downarrow}a_{\mathbf{k}'+\mathbf{q}/2,\uparrow}\nonumber\\
& &+\alpha_{b}\sum_{\mathbf{kq}} (d_{\mathbf{q}}^{\dagger}a_{-\mathbf{k}+\mathbf{q}/2,\downarrow}a_{\mathbf{k}+\mathbf{q}/2,\uparrow} + {\rm H.C.}).
\end{eqnarray}
Here, $a_{\mathbf{k}\sigma}$ and $d_{\mathbf{q}}$ are the annihilation operators for fermionic atoms and dressed molecules with mass $m$ and $2m$, respectively, and $\xi_{\mathbf{k}} \equiv \epsilon_\mathbf{k} - \mu = \mathbf{k}^2/2m-\mu$ and $\xi_{\mathbf{q}}= \epsilon_\mathbf{q}/2 - 2\mu$ are the corresponding dispersions with 2D momenta $\mathbf{k}$ and $\mathbf{q}$. The bare detuning $\delta_{b}$, bare coupling constant between atom and dressed molecule $\alpha_{b}$, and bare background interaction in the open channel $V_{b}$ should be renormalized to their corresponding physical parameters $\delta_{p}$, $\alpha_{p}$ and $V_{p}$, via a standard procedure to eliminate divergence and to recover the two-body physics at low energy~\cite{Duan2} (see details in Supplemental Material~\cite{supp}). To connect with the experiments~\cite{Moritz1}, these parameters are chosen to describe a Q2D Fermi gas of $^{6}$Li atoms near a wide Feshbach resonance at 834 G, with an effective 2D density per spin state of $n=0.8\,{\rm atoms}/{\rm \mu m}^{2}$ and an axial trapping frequency $\omega_{0}=2 \pi \times 9.2$ kHz.

We solve the effective two-channel model Hamiltonian by using the GPF theory, which is reliable at zero temperature~\cite{Hu2, Hu3, He-15}. The theory can be easily formulated with the help of the functional path-integral approach as detailed in Supplemental Material~\cite{supp}. Here, we only briefly review the key steps. By applying the Hubbard-Stratonovich transformation, we first decouple the interaction term by introducing an auxiliary pairing field and integrate out the fermionic fields $a_{\mathbf{k}\sigma}$. We then find an effective action describing the pairing field and molecule field~\cite{supp}. By taking the saddle-point solution of both fields, we obtain the mean-field thermodynamic potential,
\begin{eqnarray}
\Omega_{\rm MF}=\sum_{\mathbf{k}}\left(\xi_{\mathbf{k}}-E_{\mathbf{k}}+\frac{\Delta^{2}}{2\epsilon_{\mathbf{k}}+\omega_{z}}\right)-\frac{\Delta^{2}}{V_{\rm eff}^p},
\end{eqnarray}
where $E_{\mathbf{k}}=\sqrt{\xi_{\mathbf{k}}^{2}+\Delta^{2}}$ is the quasiparticle dispersion and $\Delta$ is the superfluid gap. $V_{\rm eff}^p=V_{p}+\frac{\alpha_{p}^{2}}{2\mu-\delta_{p}}$ is the renormalized effective interaction, obtained by replacing the bare effective interaction $V_{\rm eff}^b=V_{b}+\frac{\alpha_{b}^{2}}{2\mu-\delta_{b}}$ with the physical ones via the renormalization relation $1/{V_{\rm eff}^b}=1/{V_{\rm eff}^p}-\sum_{\bf k}\frac{1}{2\epsilon_{\bf k}+\omega_z}$~\cite{supp}. The saddle-point condition of $\Omega_{\rm MF}$ implies that the superfluid gap is determined at the mean-field level,
\begin{equation}
\frac{1}{V_{\rm eff}^p} = -\sum_{\mathbf{k}}\left(\frac{1}{2E_{\mathbf{k}}}-\frac{1}{2\epsilon_{\mathbf{k}}+\omega_{z}}\right). \label{eqn:gap}
\end{equation}
The approximate saddle-point solution can be significantly improved by the inclusion of Gaussian pair fluctuations, which gives rise to the GPF thermodynamic potential,
\begin{equation}
\Omega_{\textrm{GPF}}=\frac{1}{2}\sum_{q}\ln\left[M_{11}\left(q\right)M_{22}\left(q\right)-M_{12}\left(q\right)M_{21}\left(q\right)\right]e^{i\nu_{n}0^{+}},\label{eq:OmegaGPF}
\end{equation}
where $q=(\mathbf{q},i\nu_n)$ with the bosonic Matsubara frequency $\nu_n=2n\pi T$ ($n\in \mathbb{Z}$) and $e^{i\nu_{n}0^{+}}$ is added to ensure the convergence of the summation over $q$. The expressions of the elements of the Gaussian fluctuation matrix $M_{ij}(q)$ ($i,j=1,2$) are given by,
\begin{widetext}
\begin{eqnarray}
M_{11}(q)&=&M_{22}(-q)=\sum_{\bf k} \left(\frac{u_{{\bf k}+}^{2}u_{{\bf k}-}^{2}}{i\nu_{n}-E_{{\bf k}+}-E_{{\bf k}-}}-\frac{v_{{\bf k}+}^{2}v_{{\bf k}-}^{2}}{i\nu_{n}+E_{{\bf k}+}+E_{{\bf k}-}}+\frac{1}{2\epsilon_{\bf k}+\omega_{z}}\right)-\frac{1}{V_{\rm eff}^p(q)}, \nonumber \\
M_{12}(q)&=&M_{21}(q)=\sum_{\bf k}\left(\frac{u_{{\bf k}+}v_{{\bf k}+}u_{{\bf k}-}v_{{\bf k}-}}{i\nu_{n}+E_{{\bf k}+}+E_{{\bf k}-}}-\frac{u_{{\bf k}+}v_{{\bf k}+}u_{{\bf k}-}v_{{\bf k}-}}{i\nu_{n}-E_{{\bf k}+}-E_{{\bf k}-}}\right),
\end{eqnarray}
\end{widetext}
where  $V_{\rm eff}^p(q)=V_{p}+\frac{\alpha_{p}^{2}}{i\nu_{n}-(\mathbf{q}^{2}/4m-2\mu+\delta_{p})}$, and to simplify notation we have defined $\mathbf{k}_{\pm}=\mathbf{q}/2 \pm \mathbf{k}$. The BCS quasiparticle wavefunctions $u_{\mathbf{k}\pm}^{2}=(1+{\xi_{{\bf k}\pm}}/{E_{{\bf k}\pm}})/2$ and $v_{\mathbf{k}\pm}^{2}=(1-{\xi_{{\bf k}\pm}}/{E_{{\bf k}\pm}})/2$. With that, the number equation including Gaussian fluctuations reads
\begin{eqnarray}
\label{eqn:num}
n=-\frac{\partial\Omega_{\rm MF}}{\partial\mu}-\frac{\partial\Omega_{\rm GPF}}{\partial\mu},
\end{eqnarray}
which is solved self-consistently with the gap equation (\ref{eqn:gap}) for the chemical potential $\mu$ and superfluid gap $\Delta$.

\begin{figure}[tbp]
\includegraphics[width=1\linewidth]{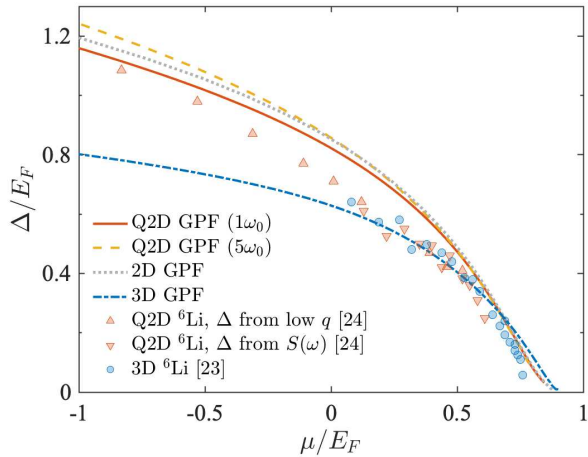}\hfill{}
\caption{\label{fig1}(Color online) Superfluid gap versus chemical potential. Results obtained via GPF approach are shown for Q2D (red solid and yellow dashed lines), 2D (gray dotted line) and 3D (blue dashed-dotted line) gases of $^{6}$Li atoms. Experimental data are extracted from recent works of 3D (blue dots)~\cite{Moritz2} and Q2D (red triangles)~\cite{Moritz1} systems.}
\end{figure}

\textit{BCS--BEC crossover.--}
The relation between the order parameter and chemical potential obtained from the GPF theory for Q2D systems are plotted in Fig.~\ref{fig1}, together with similar calculations in strictly 2D and 3D systems, and experimental data of ultracold atomic gases in Q2D~\cite{Moritz1} and 3D~\cite{Moritz2} traps. In order to compare results in different dimensions and trapping configurations, the Fermi energy $E_F$ is used as the energy unit. Notably, the results do not fall into a universal curve. In the weakly interacting regime $\mu/E_F \to 1$, theoretical calculations for all cases predict a fast decay for $\Delta$ as one would expect for a BCS superconductor. However, with increasing interaction ($\mu \to -\infty$), the order parameters in Q2D and 2D configurations are significantly elevated above the 3D case, owing to the enhancement of density of states in low dimensions. The same trend can also be observed by comparing the two Q2D systems with different axial trapping frequencies, where the result for a tight trap with $\omega_z = 5 \omega_0$ lies above the one for a loose trap with $\omega_z = 1 \omega_0$ (that corresponds to the parameters in Ref.~\cite{Moritz1}). The experimental data obtained in Q2D and 3D traps agree well with the corresponding GPF theories in both the BCS and BEC regimes. However, the deviation between theory and experiment is more drastic around unitarity $|\mu| \sim 0$, where quantum fluctuations are most significant.

\begin{figure}[tbp]
\includegraphics[width=1\linewidth]{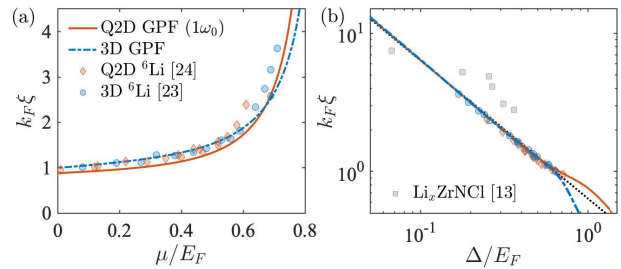}\hfill{}
\caption{\label{fig5}(Color online) Dimensionless pair size $k_{F}\xi$ as a function of (a) dimensionless chemical potential $\mu/E_{F}$ and (b) dimensionless pairing order parameter $\Delta/E_{F}$ for Q2D (red solid) and 3D (blue dashed dotted) Fermi superfluids. The black dotted line in panel (b) presents the coherence length $\xi_{p}=\frac{k_{F}}{\pi m\Delta}$. Results obtained from GPF theory are compared with experimental data for $^{6}$Li atomic gases (blue dots and red diamonds)~\cite{Moritz1, Moritz2} and Li$_x$ZrNCl (gray squares)~\cite{Nakagawa-21}.}
\end{figure}

Pair size is another fundamental physical quantity to characterize the BCS--BEC crossover, which is defined in a form independent of dimensionality
\begin{eqnarray}
\label{eqn:pairsize}
\xi^2 \equiv - \frac{\langle \varphi_{\bf k} | \nabla_{\bf k}^2 |\varphi_{\bf k}  \rangle}{\langle \varphi_{\bf k} | \varphi_{\bf k}  \rangle},
\end{eqnarray}
where $|\varphi_{\bf k} \rangle = \Delta / 2 E_{\bf k}$ is the zero temperature pair wave function. To compare with different geometries, we take the dimensionless chemical potential $\mu/E_{F}$ as a measure of interaction strength and plot dimensionless pair size $k_{F}\xi$ in Fig.~\ref{fig5}(a). Our results show quantitative consistency with the experiment data. Furthermore, the superfluid gap against pair size is plotted in Fig.~\ref{fig5}(b), where the curves for Q2D and 3D systems almost collapse onto a same straight line in the BCS regime, and start showing drastic difference in the strong interaction regime. This observation is consistent with the results shown in Fig.~\ref{fig1}. In the weak coupling regime, all systems have a same universal equation of state as expected for a BCS superconductor, and the physical properties are solely determined by the density of states at the Fermi energy. With increasing interaction, when the superfluid order parameter is comparable to the Fermi energy and the pair size reaches the order of $1/k_F$, the specific shape of dispersion will play a key role in correcting the gap. In this case, a Q2D or 2D system is more favorable for pairing as compared with a 3D system, as demonstrated both in Figs.~\ref{fig1} and \ref{fig5}. Note that in the experiment, the interaction strength is changed by tuning the scattering length via a Feshbach resonance and the chemical potential is extracted with assistance of the auxiliary filed quantum Monte Carlo (QMC) calculations~\cite{Shi}, together with the modified binding energy for a Q2D geometry given by Ref.~\cite{Levinsen}. In the Supplemental Material, we show the order parameter and pair size as a function of scattering length and compare them with the experiment data~\cite{supp}.

\begin{figure}[t!]
\includegraphics[width=1\linewidth]{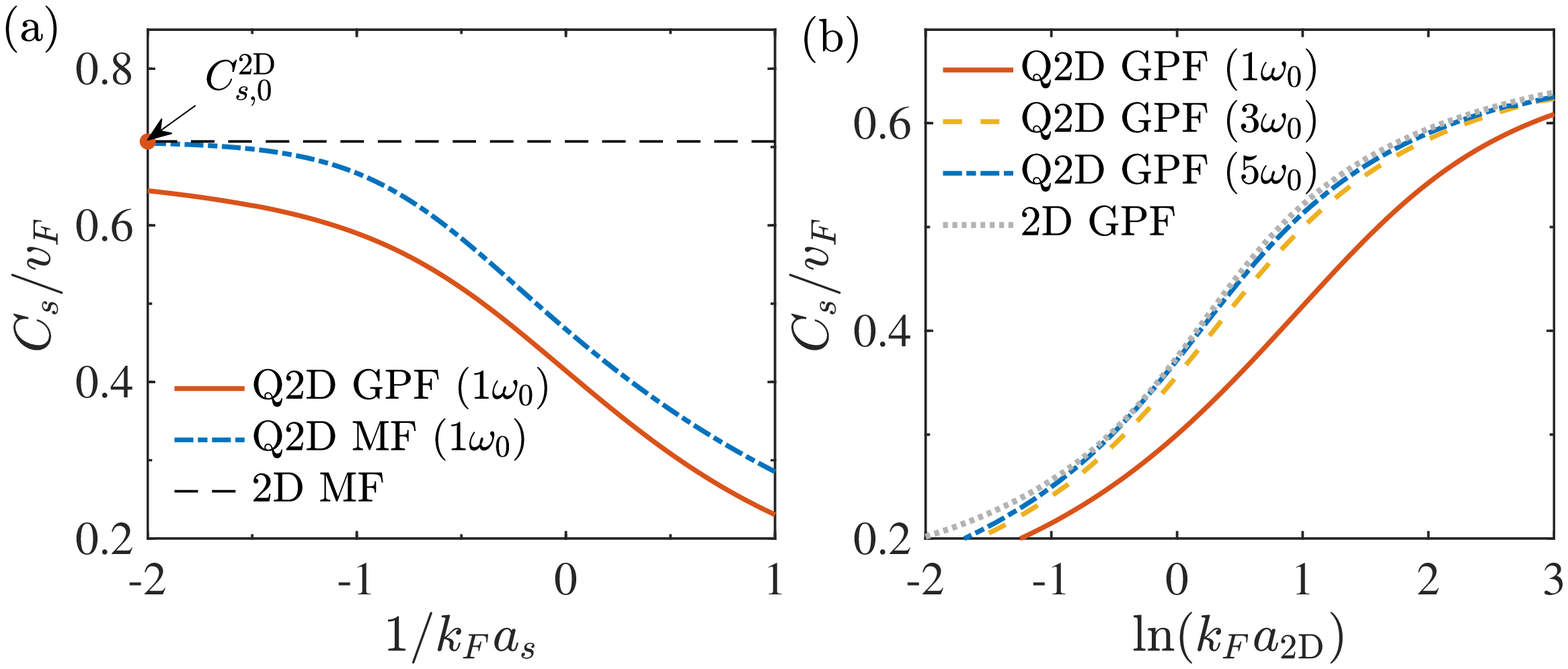}\hfill{}
\caption{\label{fig3}(Color online) (a) The sound velocity $C_{s}/v_{F}$ as a function of the 3D scattering length $a_s$. Results are obtained from Q2D GPF theory (red solid line), Q2D mean-field approximation (blue dashed-dotted line), and 2D mean-field approximation (black dashed line). The red point represents the BCS limit of $C_{s,0}^{\rm 2D}/v_{F}=1/\sqrt{2}$. The axial trapping frequency is assumed as $\omega_z = \omega_0$. (b) The sound velocity obtained from GPF theory plotted as a function of effective 2D scattering length $a_{\rm 2D}$. A pure 2D system (gray dotted line) and three Q2D configurations with $\omega_z = \omega_0$ (red solid line), $3\omega_0$ (yellow dashed line), and $5 \omega_0$ (blue dashed-dotted line) are shown for comparison.}
\end{figure}

\textit{Speed of sound and convexity of the Goldstone mode.-- }Let  us now consider the first sound velocity $C_s$, defined as the slope of the Goldstone mode in the long-wave-length limit and corresponding to the propagating speed of density waves in superfluid. Specifically, we expand the elements of the Gaussian fluctuation matrix $M(\mathbf{q}, \omega)$ to quadratic order in both $\mathbf{q}$ and $\omega$, and determine the speed of sound by requiring $\mathrm{det}M(\mathbf{q},C_{s}q)=0$. For strictly 2D systems within the mean-field approximation, the sound velocity is (incorrectly) independent on the interaction strength and reads $C_{s,0}^{\rm 2D}={v_{F}}/{\sqrt{2}}$ as shown in Fig.~\ref{fig3}(a). For Q2D systems, however, with increasing interaction, $C_{s}/v_F$ decreases from $1/\sqrt{2}$ since the excitation of high energy modes in the confined direction would suppress the propagation of density waves. Taking into account the contribution of pair flucutations, $C_{s}$ shows an obvious down-shift compared with the mean-field result, which is naturally expected since fluctuations will reduce the superfluid order parameter and consequently the sound velocity, as depicted in Fig.~\ref{fig3}(a).

To further elaborate the effect of Q2D confinement, we show in Fig.~\ref{fig3}(b) the results of $C_{s}$ for different trapping frequencies. Since the effective interaction strength is also affected by the external potential, we plot the results as a function of the dimensionless 2D interaction parameter $\ln (k_F a_{\rm 2D})$ with a 2D scattering length $a_{\rm 2D}= 2e^{-\gamma_E}/(m E_B)^{1/2}$ and the Euler constant $\gamma_E\approx 0.577$. Note that $C_{s}$ increases with a stronger confinement, and is bounded from the top by the asymptotic value in 2D systems, where the axial fluctuations are completely frozen out.

\begin{figure}[t!]
\includegraphics[width=1\linewidth]{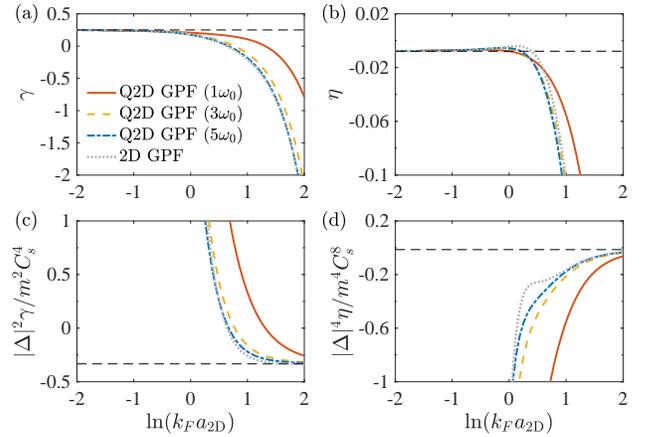}\hfill{}
\caption{\label{fig4}(Color online) Convexity parameters $\gamma$ and $\eta$ for the collective modes versus ${\rm ln}(k_{F}a_{\rm 2D})$ for 2D and Q2D systems with different trapping frequencies. A pure 2D system (gray dotted line) and three Q2D configurations with $\omega_z = \omega_0$ (red solid line), $3\omega_0$ (yellow dashed line), and $5 \omega_0$ (blue dashed-dotted line) are shown for comparison. The black dashed lines correspond to the limit values of (a) $\gamma_{\rm BEC}=1/4$, (b) $\eta_{\rm BEC}=-1/128$, (c) $\gamma_{\rm BCS} |\Delta|^2 / m^2 C_{s}^4 = -1/3$, and (d) $\eta_{\rm BCS} |\Delta|^4 / m^4 C_{s}^8 = -1/72$.}
\end{figure}

Another characteristic property of the Goldstone mode is the convexity of its dispersion relation, which is important to determine the damping mechanism~\cite{Landau,Beliaev,Kurkjian,Zou-18}. By expanding the dispersion of low-energy collective excitation up to a fifth order polynomial of $q=|\mathbf{q}|$ as $\omega_q=C_{s}q[1+\frac{\gamma}{8}(\frac{q}{mC_s})^{2}+\frac{\eta}{16}(\frac{q}{mC_s})^{4}]$, and substituting it back into the elements of the $M$ matrix, we can rewrite the condition $\mathrm{det}M(q,\omega_q)=0$ to the sixth order of $q$ and determine the convexity parameters $\gamma$ and $\eta$ from the coefficients of terms $q^{4}$ and $q^{6}$, respectively. In Fig.~\ref{fig4}, we compare the results in 2D and Q2D systems with different trapping frequencies. In the BEC regime, the convexity is convex $(\gamma>0)$, indicating that the dominant damping mechanism of collective excitation is the Landau-Beliaev two-photon to one-photon process. In the strongly interacting limit, $\gamma$ approaches a universal value of $1/4$, which is determined by the dispersion of Bogoliubov excitations in a weakly interacting Bose gas with mass $2m$ and is valid in all dimensions and trapping geometries. When the interaction is reduced, the system crosses over to the BCS regime where the damping mechanism is dominated by the Landau-Khalatnikov two-photon to two-photon process, and the convexity turns to concave ($\gamma<0$). The sign changing point of $\gamma$ moves towards the BCS side with decreasing axial trapping frequency, presenting the same trend as the first sound velocity shown in Fig.~\ref{fig3}(b).
In the BCS limit where $\Delta$ and $E_F$ are the only relevant energy scales, $\omega_{\bf q}/\Delta$ becomes a universal function of $C_s q/\Delta$. A dimensional analysis suggests that $\gamma |\Delta|^2 / m^2 C_{s}^4$ for all the 2D and Q2D cases should approach a same asymptotic value of $-1/3$, as shown in Fig.~\ref{fig4}(c). For the other parameter $\eta$, we observe qualitatively the same behavior where $\eta$ approaches a universal value of $-1/128$ in the BEC limit [Fig.~\ref{fig4}(b)], and $\eta |\Delta|^4 / m^4 C_{s}^8$ saturates to $-1/72$ for all 2D and Q2D systems [Fig.~\ref{fig4}(d)]. Our results of the convexity parameters can be tested in future experiments on Q2D Fermi systems.

\textit{Conclusions.-- }We have studied the BCS--BEC crossover in a quasi-two-dimensional quantum gas of ultracold fermionic atoms. By adopting an effective model of dressed molecule to take into account the excited degrees of freedom and using the Gaussian pair fluctuation approach, we characterize various pairing parameters at zero temperature in a wide range of crossover. We have found that the order parameter and pair size show universal relations with the chemical potential on the BCS side, but behave distinctively with increasing interaction. These results reveal the notable effect of reduced dimensionality on pairing physics, which can also be seen in the speed of first sound and convexity of Goldstone modes. This observation rules out the existence of dimensionality-independent universal relations for the equations of state, as recently suggested~\cite{Moritz1}. Our results of order parameter, chemical potential, pair size, and convexity parameters show good agreement with recents experiments of $^6$Li atomic gases\cite{Moritz1, Moritz2} and Li$_x$ZrNCl~\cite{Nakagawa-21}, and known asymptotic values in limiting cases.

\begin{acknowledgments}
This work is supported by the National Natural Science Foundation of China (Grants No.~12074428, No.~92265208), the Beijing Natural Science Foundation (Grant No.~Z180013), and the National Key R\&D Program of China (Grant No.~2018YFA0306501).
\end{acknowledgments}

\end{document}


\title{ Supplemental Material for ``BCS-BEC crossover in a quasi-two-dimensional Fermi superfluid"}
\author{Jing Zhou}
\thanks{These authors contributed equally to this work.}
\affiliation{Department of Science, Chongqing University of Posts and Telecommunications,
	Chongqing 40006, China}
\author{Tingting Shi}
\thanks{These authors contributed equally to this work.}
\affiliation{Department of Physics, Renmin University of China, Beijing 100872, China}
\author{Xia-Ji Liu}
\affiliation{Centre for Quantum Technology Theory, Swinburne University of Technology,
	Melbourne, Victoria 3122, Australia}
\author{Hui Hu}
\thanks{corresponding author:hhu@swin.edu.au}
\affiliation{Centre for Quantum Technology Theory, Swinburne University of Technology,
	Melbourne, Victoria 3122, Australia}
\author{Wei Zhang}
\thanks{corresponding author:wzhangl@ruc.edu.cn}
\affiliation{Department of Physics, Renmin University of China, Beijing 100872, China}
\address{Beijing Academy of Quantum Information Sciences, Beijing 100193, China}

\maketitle

\onecolumngrid

\setcounter{figure}{0}
\renewcommand{\thefigure}{S\arabic{figure}}

\section{The relation between 2D and 3D physical parameters}
In cold atomic gases where atoms move freely in two dimensions but are strongly confined along the axial direction by a one-dimensional harmonic trap of frequency $\omega_z$, the confined dimension still has a residual effect because of the populations of the bound molecule states in the excited levels even for an extremely strong trap. To capture this residual effect, we follow Ref.~\cite{Duan1} to convert the actual 3D Hamiltonian to an effective 2D Hamitonian of Eq.~(1) in the main text with a renormalized interaction between the atoms in the axial ground state and the dressed molecules. The 2D bare parameters $\delta_b$, $\alpha_b$ and $V_b$ in the model Hamiltonian are connected to the 2D physical parameters $\delta_p$, $\alpha_p$ and $V_p$ via the renormalization relation ($\hbar=1$)
\begin{eqnarray}
\frac{1}{V_b + \frac{\alpha_b^2}{2\mu-\delta_b}} = \frac{1}{V_{\rm eff}^p} - \sum_{\bf k}\frac{1}{2\epsilon_{\bf k}+\omega_z},
\end{eqnarray}
where $V_{\rm eff}^p=V_p+\frac{\alpha_p^2}{2\mu-\delta_p}$.

The 2D physical parameters $\delta_p$, $\alpha_p$ and $V_p$ are determined by matching the two-body bound state from the original 3D Hamiltonian, with the 3D dimensionless physical parameters $U_p=4\pi a_{\rm bg}/a_z$, $g_p^2=\mu_{\rm co}WU_p/\omega_z$, and $\nu_p=\mu_{\rm co}(B-B_0)/\omega_z$. Here, $a_{\rm bg}$ is background scattering length, $a_z=\sqrt{1/m\omega_z}$ is the characteristic length for axial motion, $\mu_{\rm co}$ is the difference in magnetic moments between the open and close channels, $W$ is the resonance width, and $B_0$ is the resonance position. To compare with the experiments of 2D $^{6}{\rm Li}$ atomic gases~\cite{Moritz1}, we will use the parameters $a_{bg}=-1405a_{0}$, $W=300$G, and $\mu_{\rm co}=2\mu_{B}$, where $a_0$ and $\mu_B$ are respectively the Bohr radius and Bohr magneton. Two functions $S_p(E)$ and $\sigma_p(E)$ turn out to be important for matching the two-body bound state and the parameter renormalization~\cite{Duan1}:
\begin{eqnarray}
S_p(E) &=& -\frac{1}{2^{5/2}\pi}\int_0^\infty dx\Big[\frac{\Gamma(x-E/2)}{\Gamma(x+1/2-E/2)} - \frac{1}{\sqrt{x}}\Big],\notag\\
\sigma_p(E) &=& \frac{{\rm ln}|E|}{2^{5/2}\pi^{3/2}}.
\end{eqnarray}
For example, the two-body binding energy $E_B$ is determined by the solution of the eigen-equation
\begin{eqnarray}
1/U_{\rm eff}^p(E_B)=S_p(E_B),
\end{eqnarray}
where $U_{\rm eff}^p(E) = U_p+\frac{g_p^2}{E-\nu_p}$.

In brief, the first matching of the bound state energy in the off-resonance BCS limit leads to~\cite{Duan1}
\begin{eqnarray}
	V_{p}^{-1}=\sqrt{2\pi}(U_{p}^{-1}-C_{p}),
\end{eqnarray}
where $C_p = {\rm lim}_{\nu_p\to\infty}[S_p(E_B)-\sigma_p(E_B)]$. Then, matching the binding energy at an arbitrary detuning yields
\begin{eqnarray}
\delta_{p} & = & E_B - \frac{\sigma_p(E_B)}{\partial_E\Big[1/U_{\rm eff}^p(E)-\big(S_p(E)-\sigma_p(E)\big)\Big]\Big|_{E=E_B}} \bigg(1-\frac{\sigma_p(E_B)}{U_p^{-1}-C_p}\bigg),\notag\\
\alpha_p^2& = & \frac{1/\sqrt{2\pi}}{\partial_E\Big[1/U_{\rm eff}^p(E)-\big(S_p(E)-\sigma_p(E)\big)\Big]\Big|_{E=E_B}}
\bigg(1-\frac{\sigma_p(E_B)}{U_p^{-1}-C_p}\bigg)^2.
\label{eq:3dpara}
\end{eqnarray}
%

\section{Gaussian pair fluctuation theory}
Adopting the functional path-integral formalism over imaginary time $\tau$ and applying the Hubbard-Stratonovich transformation in real space~\cite{Hu-20}, we decouple the interaction term by introducing an auxiliary bosonic field $\Delta(x)=\alpha_{b}\phi(x)+\widetilde{\Delta}(x)$ with $\phi(x)=\langle d(x)\rangle$, $\widetilde{\Delta}(x)=V_{b}\langle a_{\downarrow}(x)a_{\uparrow}(x)\rangle$, and $x = ({\bf x}, \tau)$ the short-hand notation. The effective action is then given in term of the fermionic Nambu spinor $A(x)=[a_{\uparrow}(x), a_{\downarrow}^{\dagger}(x)]^{T}$ as
\begin{eqnarray}
S & = & \beta\sum_{\mathbf{k}}\xi_{\mathbf{k}}+\int dxdx' A(x)^{\dagger}[-\mathcal{G}_{a}^{-1}(x,x')] A(x')-\frac{|\Delta(x)|^{2}}{V_{b}}\nonumber\\
& &+\overline{d}(x)\left(\partial_{\tau}+\widetilde{\omega}_{\bf q}-\frac{\alpha_{b}^{2}}{V_{b}}\right)d(x)+\frac{\alpha_{b}}{V_{b}}[\overline{\phi}(x)\Delta(x) + {\rm H.C.}].
\end{eqnarray}
Here, $\widetilde{\omega}_{\mathbf{q}}=\mathbf{q}^{2}/4m-2\mu+\delta_{b}$, and $\mathcal{G}_{a}(x,x')$ is the Green's function (GF) of atoms. Under the mean-field approximation $\Delta(x)=\Delta$ and $\phi(x)=\phi_{0}$, the inverse GF is given by
\begin{eqnarray}
\mathcal{G}_{a}^{-1}(x,x')=\left(
                        \begin{array}{cc}
                          -\partial_{\tau}-H_{0}^{a} & -\Delta \\
                          -\Delta & -\partial_{\tau}+H_{0}^{a} \\
                        \end{array}
                      \right)\delta_{xx'}\delta_{\tau\tau_{'}},
\end{eqnarray}
where $H_{0}^{a}={-\nabla^{2}}/{2m}-\mu$ is the kinetic energy of atoms.
By minimizing the mean-field effective action with respect to $\phi_{0}$, we obtain the zero-temperature thermodynamic potential ($\Omega_{\rm MF}=k_BT S_{\rm MF}$)
\begin{eqnarray}
\Omega_{\rm MF}=\sum_{\mathbf{k}}\left(\xi_{\mathbf{k}}-E_{\mathbf{k}}+\frac{\Delta^{2}}{2\epsilon_{\mathbf{k}}+\omega_{z}}\right)-\frac{\Delta^{2}}{V_{\rm eff}^p},
\end{eqnarray}
where $E_{\mathbf{k}}=\sqrt{\xi_{\mathbf{k}}^{2}+\Delta^{2}}$ is the quasiparticle dispersion, and $V_{\rm eff}^p=V_{p}+\frac{\alpha_{p}^{2}}{2\mu-\delta_{p}}$ is the renormalized effective interaction. Note that the bare parameters are replaced by the physical ones by the renormalization relation $1/{V_{\rm eff}^b}=1/{V_{\rm eff}^p}-\sum_{\bf k}\frac{1}{2\epsilon_{\bf k}+\omega_z}$, with $V_{\rm eff}^b=V_{b}+\frac{\alpha_{b}^{2}}{2\mu-\delta_{b}}$ the bare effective interaction. The saddle point conditions of $\Omega_{\rm MF}$ lead to the gap and number equations in the mean-field level
\begin{eqnarray}
\frac{1}{V_{\rm eff}^p} & = & -\sum_{\mathbf{k}}\left(\frac{1}{2E_{\mathbf{k}}}-\frac{1}{2\epsilon_{\mathbf{k}}+\omega_{z}}\right),
\label{eqn:gap} \\
n_{\rm MF} & = & -\frac{\partial\Omega_{\rm MF}}{\partial\mu} = 2\phi_{0}^{2}+\sum_{\mathbf{k}}\left(1-\frac{\xi_{\mathbf{k}}}{E_{\mathbf{k}}}\right),
\end{eqnarray}
where $\phi_{0}^{2}={\Delta^{2}}/{[\alpha_{p}+{V_{p}}(2\mu-\delta_{p})/{\alpha_{p}}]^{2}}$ is the population of dressed molecules.

We then introduce the fluctuation of the pairing order parameters $\Delta(x)=\Delta+\delta\psi(x)$
and $\phi(x)=\phi+\delta\phi(x)$, and calculate the contribution to the effective action up to the second order. This leads to the following Gaussian term at zero temperature
\begin{eqnarray}
S_{\rm GPF} & = & -\frac{1}{2}\sum_{q}\overline{\Phi}_{m}(q)D_{0}^{-1}(q)\Phi_{m}(q)-\frac{1}{2}\overline{\Psi}(q)\Gamma_0^{-1}(q)\Psi(q)\nonumber\\
& &+\frac{\alpha_{b}}{2V_{b}}\left[\overline{\Phi}_{m}(q)\Psi(q) + {\rm H.C.}\right],
\end{eqnarray}
where the bosonic Nambu spinors are defined as $\Phi_{m}(q)=(\delta\phi(q), \overline{\delta\phi}(-q))^{T}$ and $\Psi (q)=(\delta\psi(q), \overline{\delta\psi}(-q))^{T}$, with $q = ({\bf q}, \nu_n)$ denoting both the 2D pairing momentum ${\bf q}$ and the bosonic Matsubara frequency $\nu_n$. The inverse of the bosonic GF of dressed molecule reads
\begin{eqnarray}
D_{0}^{-1}(q)=\left(
                \begin{array}{cc}
                  i\nu_{n}-\omega_{\mathbf{q}}+\frac{\alpha_{p}^{2}}{V_{p}} & 0 \\
                  0 & -i\nu_{n}-\omega_{\mathbf{q}}+\frac{\alpha_{p}^{2}}{V_{p}} \\
                \end{array}
              \right),
\end{eqnarray}
with $\omega_{\mathbf{q}}={\bf q}^{2}/4m-2\mu+\delta_{p}$. The elements of the $2\times 2$ inverse vertex function $\Gamma_0^{-1}$ are
\begin{eqnarray}
\Gamma_{0,11}^{-1}(q) & = & \Gamma_{0,22}^{-1}(-q) =  + \sum_{k}\mathcal{G}_{11}(k)\mathcal{G}_{22}(k-q)+\frac{1}{V_b},
\nonumber \\
\Gamma_{0,12}^{-1}(q) & = & \Gamma_{0,21}^{-1}(q) = - \sum_{k}\mathcal{G}_{12}(k)\mathcal{G}_{12}(k-q),
\end{eqnarray}
where $\mathcal{G}$ is the saddle-point GF with the notation $k=({\bf k},\omega_m)$,
\begin{eqnarray}
\mathcal{G}(k)=\frac{1}{(i\omega_{m})^{2}-E_{\bf k}^{2}}\left(
  \begin{array}{cc}
    i\omega_{m}+\xi_{\mathbf{k}} & -\Delta \\
    -\Delta & i\omega_{m}-\xi_{\mathbf{k}} \\
  \end{array}
\right).
\end{eqnarray}
By integrating over the two bosonic Nambu spinors~\cite{Hu-20}, we find that the thermodynamic potential $\Omega_{\rm GPF} = k_BT S_{\rm GPF}$ contributed by quantum pair fluctuations is given by
\begin{equation}
\Omega_{\textrm{GPF}}=\frac{1}{2}\sum_{q}\ln\left[M_{11}\left(q\right)M_{22}\left(q\right)-M_{12}\left(q\right)M_{21}\left(q\right)\right]e^{i\nu_{n}0^{+}},\label{eq:OmegaGPF}
\end{equation}
where an extra factor $e^{i\nu_{n}0^{+}}$ is added to ensure the convergence of the summation over $q$. The expressions of $M_{ij}(q)$($i,j=1,2$) can be worked out explicitly and at zero temperature lead to
\begin{eqnarray}
M_{11}(q)&=&\sum_{\bf k} \left(\frac{u_{{\bf k}+}^{2}u_{{\bf k}-}^{2}}{i\nu_{n}-E_{{\bf k}+}-E_{{\bf k}-}}-\frac{v_{{\bf k}+}^{2}v_{{\bf k}-}^{2}}{i\nu_{n}+E_{{\bf k}+}+E_{{\bf k}-}}+\frac{1}{2\epsilon_{\bf k}+\omega_{z}}\right)-\frac{1}{V_{\rm eff}^p(q)}, \nonumber \\
M_{12}(q)&=&\sum_{\bf k}\left(\frac{u_{{\bf k}+}v_{{\bf k}+}u_{{\bf k}-}v_{{\bf k}-}}{i\nu_{n}+E_{{\bf k}+}+E_{{\bf k}-}}-\frac{u_{{\bf k}+}v_{{\bf k}+}u_{{\bf k}-}v_{{\bf k}-}}{i\nu_{n}-E_{{\bf k}+}-E_{{\bf k}-}}\right),
\end{eqnarray}
where  $V_{\rm eff}^p(q)=V_{p}+\frac{\alpha_{p}^{2}}{i\nu_{n}-(\mathbf{q}^{2}/4m-2\mu+\delta_{p})}$.

In the expressions above, we define $\mathbf{k}_{\pm}=\mathbf{q}/2 \pm \mathbf{k}$, and BCS superfluid parameters $u_{\mathbf{k}\pm}^{2}=(1+{\xi_{{\bf k}\pm}}/{E_{{\bf k}\pm}})/2$ and $v_{\mathbf{k}\pm}^{2}=(1-{\xi_{{\bf k}\pm}}/{E_{{\bf k}\pm}})/2$ to simplify notation, where $\xi_{{\bf k}\pm}={\bf k}_{\pm}^2/2m-\mu$ and $E_{{\bf k}\pm}=\sqrt{\xi_{{\bf k}\pm}^{2}+\Delta^{2}}$. With that, the number equation including the Gaussian fluctuation effect reads
\begin{eqnarray}
\label{eqn:num}
n=-\frac{\partial\Omega_{\rm MF}}{\partial\mu}-\frac{\partial\Omega_{\rm GPF}}{\partial\mu},
\end{eqnarray}
which is solved self-consistently with the mean-field gap equation numerically to determine the chemical potential $\mu$ and order parameter $\Delta$.

\begin{figure}[htbp]
\centering
\includegraphics[width=0.8\linewidth]{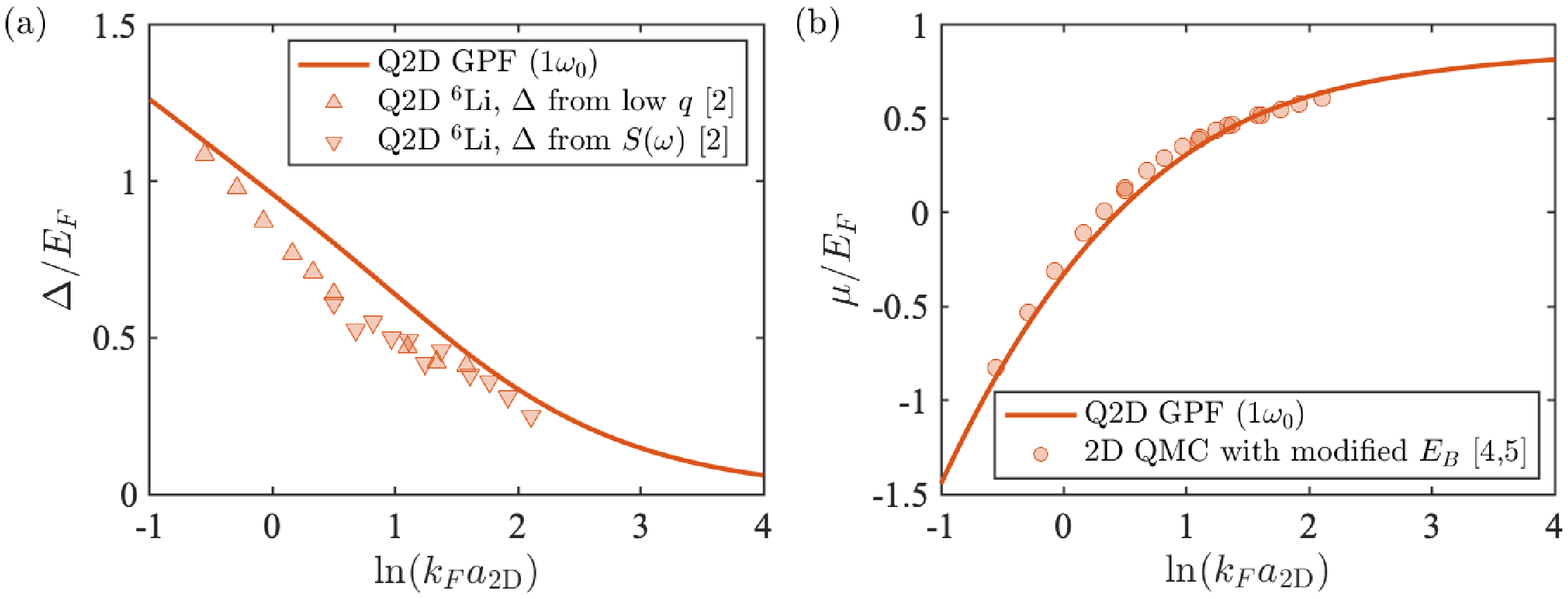}\hfill{}\caption{\label{figS1}(Color online) (a) Superfluid gap and (b) chemical potential versus the 2D interaction strength ${\rm ln}(k_Fa_{\rm 2D})$. Results obtained from Q2D GPF theory (red solid line) are compared with the experimental data from $^{6}{\rm Li}$ atomic gases (red triangles)~\cite{Moritz1} and the 2D QMC results (red dots)~\cite{Zhang} with the correction of Q2D binding energy $E_B$~\cite{Levinsen}.}
\end{figure}

\section{Superfluid gap and chemical potential}
Based on the coupled Eqs.~(3) and (6) in the main text, we determine the superfluid gap $\Delta$ and the chemical potential $\mu$ self-consistently in a wide range of crossover, where the interaction strength of pure 2D and Q2D systems can be  parameterized by the 2D dimensionless interaction parameter ${\rm ln}(k_Fa_{\rm 2D})$. We compare the numerical result of superfluid gap with the experimental data of a Q2D $^{6}{\rm Li}$ Fermi gas~\cite{Moritz1} with the same axial trapping frequency $\omega_z=1\omega_0$ in Fig.~\ref{figS1}(a). The numerical results of chemical potential are compared with 2D auxiliary field QMC calculations~\cite{Zhang} with modified Q2D binding energy $E_B$~\cite{Levinsen} in Fig.~\ref{figS1}(b).
Note that in the experiment the order parameter is measured in two ways, one is directly extracted from the integrated dynamic structure factor $S(\omega)=\int S(q,\omega)qdq$ in the BCS regime, the other is from the dynamic structure factor $S(q,\omega)$ at a small momentum transfer in the BEC regime. The chemical potential in the experiment is determined by the auxiliary field QMC calculations with the modification binding energy $E_B$ for a Q2D geometry, which is given by
\begin{eqnarray}
	\frac{a_z}{a_s} = \int_0^{+\infty}\frac{du}{\sqrt{4\pi u^3}}\left[
	1 - \frac{e^{-\frac{E_B}{\omega_z}u}}{\sqrt{(1-e^{-2u})/(2u)}}
	\right].
	\label{eq:S1}
\end{eqnarray}

We find that the experimental data of superfluid gap agree well with our theoretical prediction of a Q2D system in both the BCS and BEC regimes, while the deviation between them is more notable in the unitary region owing to significant fluctuations.
A self-consistent calculation of the chemical potential for a Q2D system is in excellent agreement with the 2D QMC results when the binding energy $E_B$ is corrected by the Q2D result~\cite{Levinsen}. This indicates that the correction plays an essential role in the connection between Q2D and purely 2D systems.

\begin{figure}[htbp]
\centering
\includegraphics[width=0.8\linewidth]{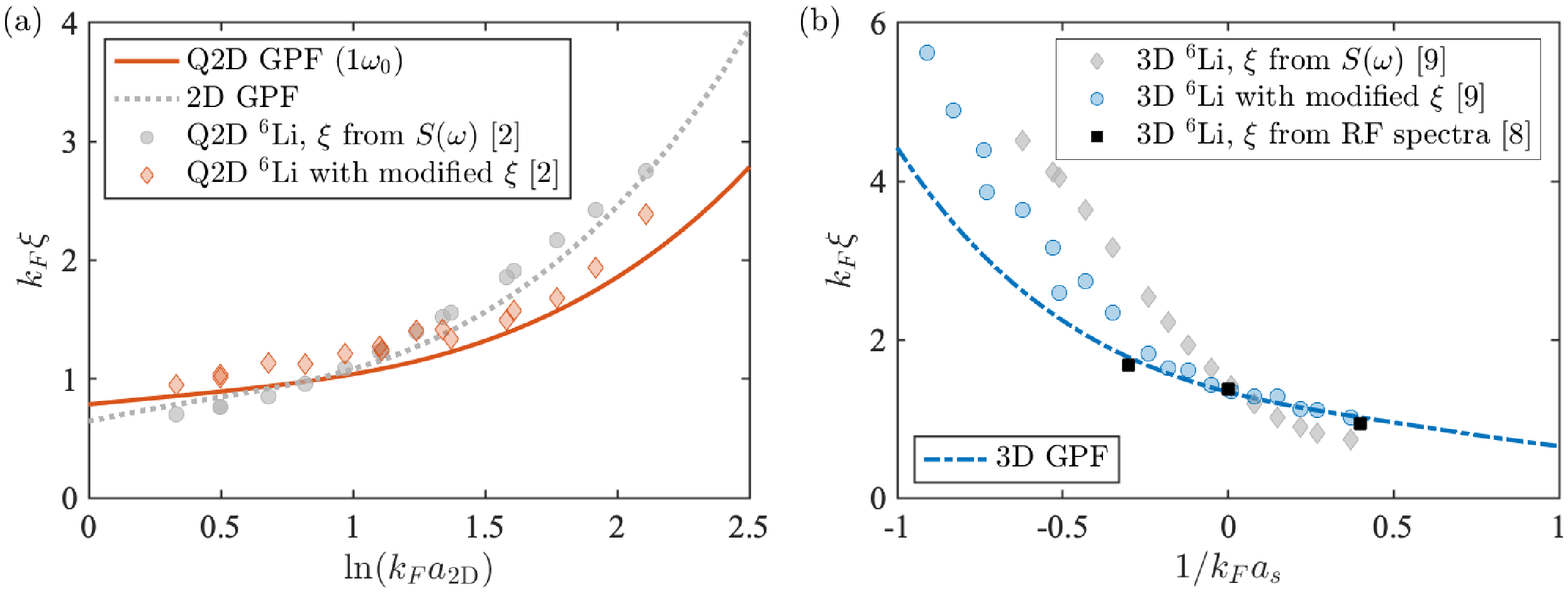}\hfill{}\caption{\label{figS2}(Color online) Dimensionless pair size $k_F\xi$ for (a) Q2D (red solid line) and 2D (gray dotted line) systems as a function of 2D interaction strength ${\rm ln}(k_Fa_{\rm 2D})$ and (b) 3D (blue dashed-dotted line) system as a function of 3D interaction strength $1/k_Fa_s$. Results obtained from GPF theory are compared with the experiment data obtained from the onset momentum $k_o$ of the pair breaking continuum for Q2D (gray dots)~\cite{Moritz1} and 3D~\cite{Moritz2} (gray diamonds) gases of $^{6}{\rm Li}$ atoms. The red diamonds and blue dots represent the Q2D and 3D pair sizes calculated via Eq.~(\ref{eq:xi2d}) and Eq.~(\ref{eq:xi3d}) using the superfluid gap and the chemical potential obtained from Ref.~\cite{Moritz1}. The black squares show the results from a pair size measurement implemented in 3D $^{6}{\rm Li}$ atomic gases using RF spectra~\cite{Ketterle}.}
\end{figure}

\section{pair size}
To estimate the pair size of Fermi gases with different dimensionalities, we follow the definition of Eq.~(7) as stated in the main text. We plot the numerical results of $k_F\xi$ for Q2D (red solid line) and 2D (gray dotted line) systems by tuning the conventional 2D interaction parameter ${\rm ln}(k_Fa_{\rm 2D})$ in Fig.~\ref{figS2}(a), and for 3D (blue dashed-dotted line) system by varying the 3D scattering length $1/k_Fa_s$ in Fig.~\ref{figS2} (b). Analytically, the pair size can be obtained as
\begin{eqnarray}
\xi^{2}=\frac{1}{4m\Delta}\bigg[\frac{\mu}{\Delta} + \frac{\mu^2+2\Delta^2}{\mu^2+\Delta^2}\Big(\frac{\pi}{2}+{\rm arctan}\frac{\mu}{\Delta}\Big)^{-1}\bigg],
\label{eq:xi2d}
\end{eqnarray}
in two dimensions~\cite{Mohit} and
\begin{eqnarray}
\xi^{2}=\frac{1}{16m\Delta^2}\bigg[2\mu + \frac{5\Delta^2+2\mu^2}{\sqrt{\mu^2+\Delta^2}}\bigg],
\label{eq:xi3d}
\end{eqnarray}
in three dimensions~\cite{Marini}.

We note that the pair size of Q2D $^{6}{\rm Li}$ atomic gases is not directly measured in experiments~\cite{Moritz1}. Instead, one determines the onset momentum $k_o$ of the pair breaking continuum from the intersection point of a bilinear fit of the dynamic structure factor at a transferred energy of $\omega=2\Delta$, which is related to pair size via $k_o=\alpha/\xi$. The prefactor $\alpha$ is evaluated by comparing the experimental results of $1/k_o$ to theoretical predictions of the pair size of 2D or 3D Fermi gases. The raw data of pair size in Q2D~\cite{Moritz1} and 3D~\cite{Moritz2} $^{6}{\rm Li}$ Fermi gases are shown in Fig.~\ref{figS2} by gray dots and gray diamonds, respectively. We find that the raw experimental data fit better with the theoretical predictions of a 2D Fermi gas than a Q2D system. However, by substituting the superfluid gap and chemical potential measured in experiment~\cite{Moritz1} to Eq.~(\ref{eq:xi2d}) and Eq.~(\ref{eq:xi3d}), we can obtain a modified pair size. These modified results for Q2D (red diamonds) and 3D (blue dots) $^6$Li gases agree well with our Q2D and 3D predictions from GPF theory. In addition, the black squares represent the pair size of 3D cold atomic gases of $^{6}{\rm Li}$ performed in Ref.~\cite{Ketterle} obtained from radio-frequency spectra, also showing a good match to our prediction.

\begin{figure}[htbp]
\centering
\includegraphics[width=0.8\linewidth]{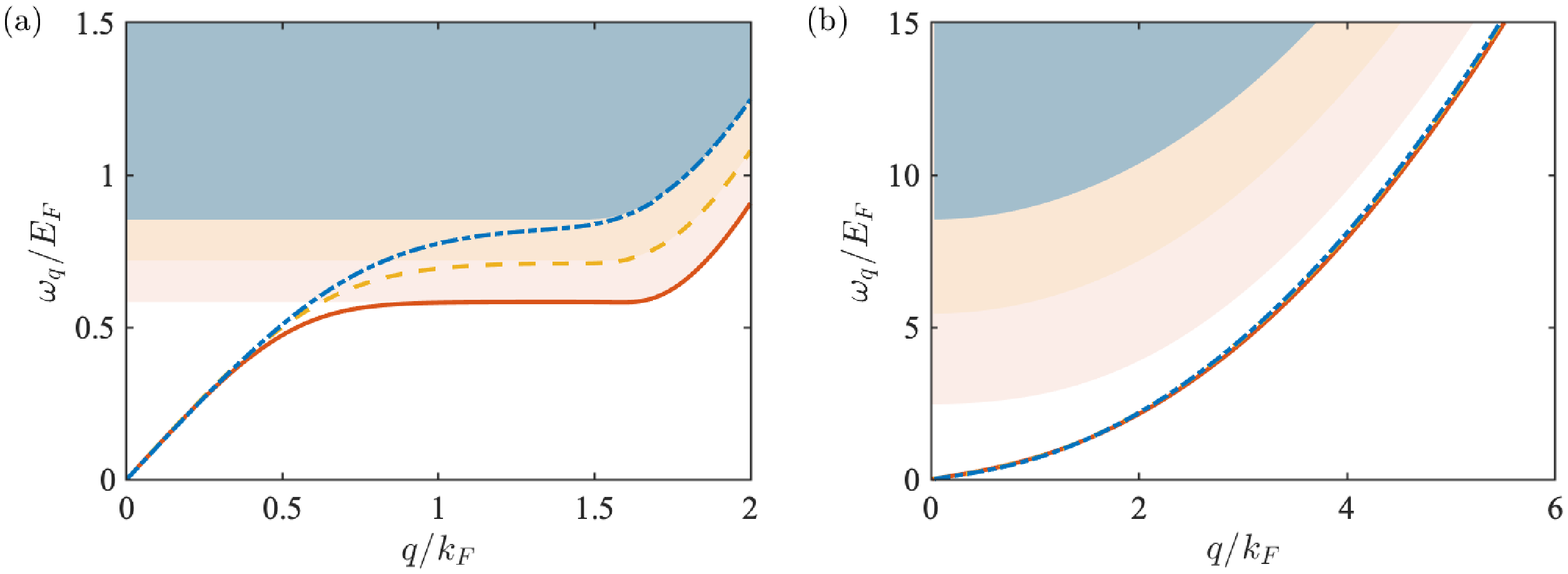}\hfill{}\caption{\label{figS3}(Color online) Collective mode spectrum $\omega_q$ versus $q$ for Q2D systems with different trapping frequencies and interactions of (a) $1/k_Fa_s=-1$ and (b) $1/k_Fa_s=1$. Three Q2D configurations with $\omega_z = \omega_0$ (red solid line), $3\omega_0$ (yellow dashed line), and $5 \omega_0$ (blue dashed-dotted line) are shown for comparison. The shades represent the continuum of pair breaking excitations.}
\end{figure}

\section{Collective mode}

The low-energy collective mode spectrum $\omega_q$ is well-known as the Goldstone mode arising from the broken of $U(1)$ symmetry, which is determined by the pole of the inverse Gaussian fluctuation boson propagator $M$, i.e. the solution of ${\rm det}[M(q,\omega_q)]=0$. The two-particle continuum begins at $E_{c}(q)={\rm min}_{\bf k}(E_{+}+E_{-})$, corresponding to the single particle excitation from the pair breaking threshold.

In Fig.~\ref{figS3}, we present the Goldstone mode by lines and pair breaking continuum by shades with different trapping frequencies on the BCS side with the interaction strength $1/k_Fa_s=-1$ [Fig.~\ref{figS3}(a)], and on the BEC side with the interaction strength $1/k_Fa_s=1$ [Fig.~\ref{figS3}(b)], where the trapping frequencies are $1\omega_0$ by red solid lines and red shades, $3\omega_0$ by yellow dashed lines and yellow shades, $5\omega_0$ by blue dashed-dotted lines and blue shades. In the long-wavelength limit, the dispersions exhibit linear shape for all configurations, the slope of which determines the sound velocity. We find that $\omega_q$ is always below the two-particle continuum, indicating the existence of a two-body bound state in a Q2D system over a wide range of crossover.

As shown in Fig.~\ref{figS3}(a), in the BCS regime, we observe a wide pair breaking continuum and the pairs are weakly bound. When the transferred energy is strong enough to break a Cooper pair, the atoms will be excited to the continuum state with a sharp energy threshold of $2\Delta$. In particular, for large transferred momentum, the Goldstone mode touches the pair breaking continuum and bends down instead of keeping a linear slope. However, in the BEC regime where the atom pairs condense as tightly-bound molecules, the continuum is notably lifted until moves out of the spectra, as shown in Fig.~\ref{figS3}(b).
